\documentclass[12pt]{article}
\usepackage{epsfig}
\usepackage{float}
\usepackage{amssymb,amsmath}
\newcommand{\be}{\begin{equation}}
\newcommand{\ee}{\end{equation}}
\newcommand{\bea}{\begin{eqnarray}}
\newcommand{\eea}{\end{eqnarray}}
\begin{document}

\begin{center}
{\bf RECOILLESS RESONANCE ABSORPTION OF TRITIUM ANTINEUTRINOS AND
TIME-ENERGY UNCERTAINTY RELATION}
\end{center}
\begin{center}
S. M. Bilenky
\end{center}

\begin{center}
{\em  Joint Institute for Nuclear Research, Dubna, R-141980,
Russia\\}
\end{center}
\begin{abstract}
We discuss   neutrino oscillations in an experiment with M\"ossbauer
recoilless resonance absorbtion of tritium antineutrinos, proposed
recently by Raghavan. We demonstrate that small energy uncertainty
of  antineutrinos which ensures a large resonance absorption cross
section is in a conflict with the energy uncertainty which,
according to the time-energy uncertainty relation, is necessary for
neutrino oscillations to happen. The search for neutrino
oscillations in the M\"ossbauer neutrino experiment would be an
important test of the applicability of the time-energy uncertainty
relation to a newly discovered  interference phenomenon.
\end{abstract}

\section{Introduction}
Uncertainty relations play an important role in the quantum theory.
They are based on fundamental general properties of the theory and
manifest the nature of it. There are two different types of the
uncertainty relations in the quantum theory: Heisenberg uncertainty
relations and time-energy uncertainty relations.

The  Heisenberg uncertainty relations are based on commutation
relations for hermitian operators, which correspond to physical
quantities. Let us consider two hermitian operators $A$ and $B$.
From the Cauchy-Schwarz inequality we have (see, for example,
\cite{Messiah})
\begin{equation}\label{Couchy}
 \Delta_{a}A~\Delta_{a}B \geq \frac{1}{2}|\langle
 a|[A,B]|a\rangle|~,
\end{equation}
Here $|a\rangle$ is some state and
\begin{equation}\label{standarddev}
 \Delta_{a}A=\sqrt{\langle a|(A-\langle a|A| a\rangle)^{2}|a\rangle                }
\end{equation}
is the standard deviation of  $A$ in the state $|a\rangle$. If
operators $A$ and $B$ satisfy the commutation relation $[A,B]=iC$,
where $C$ is a hermitian operator, than  we have the uncertainty
relation
\begin{equation}\label{Heiuncertainty1}
\Delta_{a}A~\Delta_{a}B \geq \frac{1}{2}|\langle
 a|C|a\rangle|~.
\end{equation}
For canonically conjugated quantities the right-handed parts of the
Heisenberg uncertainty relations do not depend on the state
$|a\rangle$. For example, for operators of momentum $p$ and
coordinate $q$, which satisfy the commutation relation $[p,q]=1/i$,
from (\ref{Couchy}) we obtain the standard uncertainty relation
\begin{equation}\label{Heiuncertainty2}
\Delta_{a}p~\Delta_{a}q \geq \frac{1}{2}
\end{equation}
The Heisenberg uncertainty relations are applicable to the states at
the fixed time $t$. They mean in particular  that physical
quantities whose operators do not commute can not have
simultaneously definite values.

Time-energy uncertainty relation have a completely different
character. It was a subject of intensive discussions and controversy
from the early years of the quantum theory . In the literature exist
different time-energy uncertainty relations with different meaning
of quantities which enter into them (see, for example, review
\cite{Busch07}).

The time energy uncertainty relation is based on the fact that
dynamics of a quantum system is determined by the Hamiltonian. The
most direct and general derivation of the time-energy uncertainty
relation was given by Mandelstam and Tamm \cite{TammMand45}.

Let us consider the evolution equation for any operator $O(t)$ in
the Heisenberg representation. We have
\begin{equation}\label{evolution}
    -i\frac{\partial ~O(t)}{\partial t}=[H, O(t)]~,
\end{equation}
where $H$ is the total Hamiltonian (which does not depend on time).
From (\ref{Couchy}) and (\ref{evolution}) we find
\begin{equation}\label{timeenergy1}
    \Delta_{a}E~\Delta_{a}O(t) \geq \frac{1}{2}
   | \frac{\partial }{\partial t}\langle a|O(t)|a\rangle|~.
\end{equation}

We can rewrite the inequality (\ref{timeenergy1}) in the form of the
time-energy uncertainty relation
\begin{equation}\label{timeenergy2}
    \Delta_{a}E~\Delta_{a}t \geq \frac{1}{2}~.
\end{equation}
Here
\begin{equation}\label{deltat}
\Delta_{a}t=\frac{\Delta_{a}O(t) }{| \frac{\partial }{\partial
t}\langle a|O(t)|a\rangle|}~.
\end{equation}
We  assumed that the derivative $\frac{\partial }{\partial t}\langle
a|O(t)|a\rangle$ is different from zero ($|a\rangle$ is a non
stationary state). The quantity $\Delta_{a}t$ has a dimension of
time. It depends on the state $|a\rangle$ and  operator $O(t)$.
Different systems were considered in \cite{TammMand45,Busch07}.

It follows from (\ref{deltat}) that $\Delta_{a}t $ is the time
interval which is necessary for the average value $\langle
a|O(t)|a\rangle$ to be changed by one standard deviation
$\Delta_{a}O(t)$. In other words $\Delta_{a}t $  characterizes the
time interval during which the state of the system significantly
varies.

 Neutrino oscillations is a non stationary phenomenon. This
was demonstrated by the recent accelerator K2K \cite{K2K} and MINOS
\cite{Minos} neutrino oscillation experiments in which  time of
neutrino production and neutrino detection was measured. The only
parameter which characterizes evolution of a neutrino state in the
case of neutrino oscillations is period of oscillations (or
oscillation length). In the simplest case of two neutrinos, the
period of oscillations is given by the expression\footnote{This
expression follows from the standard theory of neutrino oscillations
(see, for example, \cite{BGG}). Up to the factor $4\pi$ it can be
obtained, however, from general considerations. In fact, $E$ in the
numerator is determined by the Lorenz boost and $\Delta m^{2}$ in
the denominator follows from dimensional reasons and the
requirement: $t_{osc}\to \infty$ at $m_{2}\to m_{1}$.}
\begin{equation}\label{period}
    t_{osc}=4\pi \frac{E}{\Delta m^{2}}~,
\end{equation}
where $E$ is the neutrino energy and $\Delta m^{2}= m_{2}^{2}-
m_{1}^{2}$.

It is natural to expect that in the case of the neutrino
oscillations $\Delta t $ in the time-energy uncertainty relation
(\ref{timeenergy2}) is given by the period of oscillations
\begin{equation}\label{charactertime}
\Delta t \simeq t_{osc}~.
\end{equation}
As we will discuss in the next section accelerator neutrino
oscillation experiments confirm this expectation.

\section{Neutrino oscillations is non stationary phenomenon}

One of the most important recent discovery in the particle physics
was the discovery of neutrino oscillations \cite{SK,SNO,
Kamland,Cl,Gallex,Sage,SKsol,K2K,Minos}. All  existing at present
data are in agreement with the assumption that the fields of the
flavor neutrinos $\nu_{lL}(x)$ $(l=e,\mu,\tau)$ are mixtures of the
left-handed components of the three massive neutrino fields (see
reviews \cite{BGG,Conca})
\begin{equation}\label{3numixing}
\nu_{lL}(x)=\sum^{3}_{i=1}U_{li}\nu_{iL}(x)~.
\end{equation}
Here $U$ is the PMNS \cite{Pont57,MNS} neutrino mixing matrix  and
$\nu_{i}(x)$ is the field of neutrino with mass $m_{i}$.

In the case of the three-neutrino mixing the probabilities of the
transition between different flavor neutrinos depend on six
parameters: two mass-squared differences $\Delta m_{12}^{2}$ and
$\Delta m_{23}^{2}$, three mixing angles $\theta_{12}$,
$\theta_{23}$ and $\theta_{13}$ and CP-phase $\delta$. However, two
parameters are small: $\frac{\Delta m_{12}^{2}}{\Delta
m_{23}^{2}}\simeq 3\cdot 10^{-2}$ and $\sin^{2}2\theta_{13}\leq
5\cdot 10^{-2}$. If we neglect contribution of the small parameters
to the transition probabilities, two-neutrino
$\nu_{\mu}\leftrightarrows\nu_{\tau} $ oscillations take place in
the atmospheric-LBL region of the values of the parameter
$\frac{L}{E}$ ($L$ is the distance between neutrino production and
detection points and $E$ is the neutrino energy). For the
probability of $\nu_{\mu}$ ($\bar\nu_{\mu}$) to survive we have in
this case (see review \cite{BGG})
\begin{equation}\label{2nuprbability}
{\mathrm P}(\nu_\mu \to \nu_\mu) = {\mathrm P}(\bar\nu_\mu \to
\bar\nu_\mu)\simeq 1 - \frac {1}{2}~  \sin^{2}  2\theta_{23}~ (1 -
\cos \Delta m_{23}^{2} \frac {L}{2E})\,.
\end{equation}
In the reactor KamLAND region $\bar\nu_{e}\rightleftarrows
\bar\nu_{\mu,\tau}$ oscillations take place. The $\bar\nu_{e}$
survival probability has two-neutrino form
\begin{equation}\label{2antinuprab}
 {\mathrm P}(\bar\nu_e \to \bar\nu_e)= 1 - \frac {1}{2}~
\sin^{2}  2\theta_{12}~ (1 - \cos \Delta m_{12}^{2} \frac
{L}{2E})\,.
\end{equation}
Finally, the  probability of the solar $\nu_{e}$ to survive  is
given by the standard two-neutrino matter expression which depends
on $\Delta m_{12}^{2}$,  $\sin^{2}\theta_{12}$ and the density of
electrons in the sun.

From the analysis of the data of the Super-Kamiokande atmospheric
neutrino experiment it was found \cite{SK}
\begin{equation}\label{SK}
 1.5\cdot 10^{-3}\leq \Delta m^{2}_{23} \leq 3.4\cdot
10^{-3}~~\rm{eV}^{2},~~\sin^{2}2\theta_{23}>0.92~.
\end{equation}

From the global analysis of the data of the KamLAND and solar
neutrino experiments the following values of the parameters $\Delta
m^{2}_{12}$ and $\tan^{2}\theta_{12}$ were obtained \cite{Kamland}
\begin{equation}\label{Kamland}
\Delta m^{2}_{12} =
7.9^{+0.6}_{-0.5}~10^{-5}~\rm{eV}^{2},~~\tan^{2}\theta_{12}=0.40^{+0.10}_{-0.07}~.
\end{equation}
Finally, from analysis of the data of the reactor CHOOZ
experiment\cite{Chooz} for the  parameter $\sin^{2}\theta_{13}$ the
following upper bound was found
\begin{equation}\label{Chooz}
\sin^{2}\theta_{13}\leq 5\cdot 10^{-2}~.
\end{equation}
An important step in the study of the neutrino oscillations was the
confirmation of the results of the SK atmospheric neutrino
experiment \cite{SK} by the   long baseline K2K \cite{K2K} and MINOS
\cite{Minos} accelerator neutrino oscillation experiments. From the
analysis of data of the MINOS experiment for the parameters $\Delta
m^{2}_{23}$ and $\sin^{2}2\theta_{23}$  the following values were
obtained \cite{Minos}
\begin{equation}\label{Minos}
\Delta m^{2}_{23}= 2.74^{+0.44}_{-0.26}~~ 10^{-3}\rm{eV}^{2},~~~~
\sin^{2}2\theta_{23}>0.87~.
\end{equation}
The values (\ref{Minos}) are in agreement with (\ref{SK}).

The experiments K2K and MINOS are also  important from the point of
view of the understanding of the origin of the neutrino oscillations
(see \cite{BilMat06},\cite{BilFeilPotz07}):
 in these experiments
{\em the time of neutrino production and neutrino detection was
measured} for the first time.

Let us consider as an example the K2K experiment. In this experiment
neutrinos are produced in $1.1 ~\mu s $ spills. Protons are
extracted from the accelerator every 2.2 s. Let us denote the time
of the neutrino production at the KEK accelerator $t_{KEK}$ and the
time of the neutrino detection in SK detector $t_{SK}$. Neutrino
events which satisfy the criteria
\begin{equation}\label{}
    -0.2 \leq \left((t_{SK}-t_{KEK})-L/c\right)\leq 1.3~ \mu s~,
\end{equation}
where selected in the experiment.

In the K2K experiment the effect of neutrino oscillations was
observed. This means that during the time interval $\Delta
t=t_{SK}-t_{KEK}$  neutrino state is significantly changed (the
initial $\nu_{\mu}$-state is  transferred into a superposition of
$\nu_{\mu}$ and $\nu_{\tau}$ states). The distance $L$ in the
experiment is about 250 km and $\Delta t \simeq 0.8 \cdot 10^{3}~\mu
s$. This time is comparable with the period of oscillations  driven
by $\Delta m^{2}_{23}$ which in the K2K experiment is approximately
equal to $ 3.3\cdot 10^{3}~ \mu s$. Thus, in the case of the
neutrino oscillations $\Delta t$ in the time-energy uncertainty
relation (\ref{timeenergy2}) is of the order of the period of
oscillations.

\section{Recoilless creation and resonance absorption of  tritium
antineutrinos}

In \cite{Raghavan}  it  was proposed to detect  the tritium
$\bar\nu_{e}$ with energy $\simeq$ 18.6 KeV in the recoilless
M\"ossbauer  transitions
\begin{equation}\label{mosstransitions}
 ^{3}\rm{H}\to ^{3}\rm{He}+\bar\nu_{e},\quad \bar\nu_{e}+
^{3}\rm{He}=^{3}\rm{H}.
\end{equation}
It was estimated in \cite{Raghavan}  that the relative uncertainty
of the energy of the  antineutrinos produced in
(\ref{mosstransitions}) is of the order
\begin{equation}\label{energyuncert}
\frac{\Delta E }{E}\simeq 4.5 \cdot 10^{-16}~.
\end{equation}
With such an uncertainty it was estimated that  the cross section of
the recoilless resonance absorption of  antineutrinos in the process
$\bar\nu_{e}+ ^{3}\rm{He}=^{3}\rm{H}$ is equal to
\begin{equation}\label{crosssection}
\sigma_{R}\simeq 5\cdot 10^{-32}\rm{cm}^{2}
\end{equation}
Such a value is about nine orders of the magnitude larger than the
normal neutrino cross section.

For the tritium antineutrino with the energy $\simeq$ 18.6 KeV the
length of the oscillations driven by $\Delta m^{2}_{23}$ is given by
\begin{equation}\label{osclength}
L^{(23)}_{osc}\simeq 2.5 \frac{E(\rm{MeV})}{\Delta
m^{2}_{23}(\rm{eV^{2})}}~m \simeq 18.6 ~m
\end{equation}
It was proposed in \cite{Raghavan} to search for neutrino
oscillations in the M\"ossbauer neutrino experiment. Such
measurement would allow to determine the parameter
$\sin^{2}\theta_{13}$ (or to improve CHOOZ bound (\ref{Chooz})).
From (\ref{osclength}) follows that the baseline of such an
experiment is about 10 m.

Let us discuss possibilities of neutrino oscillations in the
M\"ossbauer neutrino experiment from the point of view of the
time-energy uncertainty relation. In order that neutrino
oscillations driven by the "large" atmospheric $\Delta m^{2}_{23}$
take place the following condition must be satisfied
\begin{equation}\label{inequality}
\frac{\Delta E }{E}\gtrsim \frac{1}{4\pi}~\frac{\Delta
m^{2}_{23}}{E^{2}}\simeq 5.8\cdot 10^{-13}~.
\end{equation}
From (\ref{inequality}) we conclude that neutrino oscillations
driven by $\Delta m^{2}_{23}$ can not be observed in the neutrino
experiment with energy uncertainty given by (\ref{energyuncert}).

We will discuss now neutrino oscillations driven by the small
solar-KamLAND neutrino mass-squared difference $\Delta m^{2}_{12}$
given by (\ref{Kamland}) in the M\"ossbauer  neutrino experiment.
The oscillation length for the tritium neutrinos will be in this
case about 30 times larger than $L^{(23)}_{osc}$. Thus, the baseline
of the experiment must be about 300 meters. This make such an
experiment very difficult. Let us see, however, whether the
experiment is possible from the point of view of the time-energy
uncertainty relation. The energy uncertainty in this case must
satisfy the inequality
\begin{equation}\label{inequality1}
\frac{\Delta E }{E}\gtrsim \frac{1}{4\pi}~\frac{\Delta
m^{2}_{12}}{E^{2}}\simeq 1.9\cdot 10^{-14}~,
\end{equation}
Thus, in the case if $\frac{\Delta E }{E}$ is given by
(\ref{energyuncert}) neutrino oscillations driven by the small
$\Delta m^{2}_{12}$ are also impossible.

It was stressed, however, in \cite{Potzel} that due to impurities,
lattice defects and other effects, which were not taken into account
in \cite{Raghavan}, the real value for $\frac{\Delta E }{E}$ can be
about two order of magnitude larger than (\ref{energyuncert}). In
this case inequality (\ref{inequality1}) could be satisfied.
However, resonance cross section will be about four order of
magnitude smaller than (\ref{crosssection}).

In conclusion, the very small energy uncertainty of antineutrinos
produced in the two-body recoilless tritium decay $^{3}\rm{H}\to
^{3}\rm{He}+\bar\nu_{e}$ which provides a very large resonance cross
section of the antineutrino absorbtion in the recoilless transition
$\bar\nu_{e}+ \rm ^{3}{He}=\rm{ ^{3}H}$  is in a conflict with the
energy uncertainty which, according to the time-energy uncertainty
relation, is necessary for neutrino oscillations to happen. Thus,
the M\"ossbauer neutrino oscillation experiment could be an
important tool for the test of the fundamental time-energy
uncertainty relation in a newly discovered interference phenomenon.
Such test can not be performed in usual neutrino oscillation
experiments (see \cite{BilFeilPotz07}).

It is a pleasure for me to thank W. Potzel, F. von Feilitzsch,
H.C.S.Lam  and Wei Liao for interesting discussions. I acknowledge
ILIAS program for support and the TRIUMF theory department for the
hospitality.

\end{document}